\documentclass[12pt]{article}
\usepackage[margin=0.7in]{geometry}
\usepackage{float}
\usepackage{amsmath}
\usepackage{amssymb}
\usepackage{graphicx}
\usepackage{cite}
\usepackage{color}
\usepackage{dcolumn}
\usepackage{bm}
\usepackage{newfloat}
\DeclareFloatingEnvironment[name={SI Figure}]{suppfigure}
\DeclareFloatingEnvironment[name={SI Table}]{supptable}
\DeclareFloatingEnvironment[name={SI equation}]{suppequation}

\begin{document}
\begin{flushleft}
{\Large
\textbf{Interaction patterns in diabetes mellitus II network: An RMT relation}
}
\\
Sarika Jalan$^{\ast,1,2}$
Aparna Rai$^{1}$, 
Amit Kumar Pawar$^{2}$, 
\\
\bf{1} Centre for Bio-Science and Bio-Medical Engineering, Indian Institute of Technology Indore, M-Block, IET-DAVV Campus,
Khandwa Road, Indore, Madhya Pradesh, 452017, India
\\
\bf{2} Complex Systems Lab, Discipline of Physics,  Indian Institute of Technology Indore, M-Block, IET-DAVV Campus,
Khandwa Road, Indore, Madhya Pradesh, 452017, India
\\
$\ast$ E-mail: sarikajalan9@gmail.com
\end{flushleft}

\section*{Abstract}

Diabetes mellitus type II affects around 8 percent of the total adult population in the world.
It is the fifth leading cause of death in high income countries and an epidemic in developing countries.
We analyze protein-protein interaction data of the pancreatic cells for normal and disease states.
The analysis exhibits overall structural similarities in the normal and disease networks.
The important differences are revealed through specific interaction patterns and eigenvector analyses.
The top contributing nodes from localized eigenvectors as well as those being part of specific interaction patterns 
turn out to be significant for the occurrence of the disease. 
The analysis provides a direction for 
further development of novel drugs and therapies in curing the disease by targeting specific
patterns instead of a single node.

\section*{Introduction}

Diabetes mellitus type II (DM-II) is a metabolic disorder which is characterized by resistance to 
insulin and abnormalities in the hepatic cells in terms of glucose production \cite{fujimoto2000}.
It has been known to be a complex and multi factorial disease \cite{mclyntyre} as it affects various 
other organ systems in the body like cardiovascular system, kidneys, lungs, eyes etc \cite{MTluna2005,Kasuga}. 
The WHO statistics reveals that DM-II has affected nearly 325 million people till 2013 out of which
$46\%$ cases are undiagnosed and the disease show rising trend world-wide \cite{who-2011,who,19}. 
Diabetes is known to cause 5.1 million death in 2013 and statistical data reveals that
every six seconds a person dies from diabetes further probing that
it is the fifth leading cause of death in high income countries and an epidemic in 
developing countries \cite{IDF,zimmet}. It is one of the most important non-communicable 
diseases of the 21st century accounted both in terms of mortality and morbidity \cite{Science_2014}.
Indecent lifestyle, increased urbanization and unhealthy behaviors are some of the major causes behind the disease \cite{lin}.
Recent biological studies indicate that significant increase in population having obesity 
has in turn become a major reason for higher DM-II cases \cite{dean}.
Some genome wide studies reveal genetic basis of DM-II by finding some important proteins 
which are heritable and cause DM-II further indicating that dysfunctioning of beta cells is another
reason for the occurrence of the disease apart from insulin resistance \cite{christian}.
This disease is likely to increase the burden of chronic diabetic 
complications worldwide in the near future \cite{rose,dabelea,kita}.
It has been found that treatment of people with diabetes account for up
to 15$\%$ of international health care budgets \cite{ada,idf_1996}. 
According to a recent study, a person with this complex disease spends on 
an average more than $\$85,000$ in the treatment and its complications over 
entire lifetime and trillion dollars are spent on health and diseases worldwide 
\cite{idf_1996}. An early detection may help in cure to the people with this disease. 
But the research pertaining to development of novel treatments and cure methods is very exhaustive \cite{IDF}. 
Even after enormous investment in pharmacological research and clinical trials in the past decades, 
there has not been proportionate advancement in clinical results. All these demand for a new perspective in disease research, 
which we provide here using combined framework of random matrix theory and network theory. 
This framework predicts important structural patterns crucial for disease is very time efficient and cost effective.

Previously, a network analysis of PPI netwok of DM-II reveals potential drug targets on the basis of 
candidate genes involved in DM-II \cite{bagler}.
Few network studies done also reveal some important pathways associated with insulin signaling 
using high-throughput micro-array datasets \cite{manway,Rasche,voight}.   
DM-II investigated using systems biology approach highlights the occurrence of the disease.
These studies are based on a handful of 
disease-associated proteins, while our investigation for the first time performs extensive 
analysis of the whole pancreatic cell proteins.
We investigate DM-II at the proteomic level using network biology along with the random matrix theory (RMT). 
This mathematical theory has been formulated six decades ago in order to understand
complex nucleons interactions \cite{Wigner}. 
Later on, the theory has shown its remarkable success in understanding various complex systems 
ranging from quantum chaos to galaxy \cite{rev_rmt,Fossion,complex_net,complex_net1,complex_net2}.
In the current work, we construct the protein-protein interaction (PPI) network of DM-II by analyzing normal and disease 
states of pancreatic cells and investigate their structural properties, which is further compared with the properties of random networks.
The analysis reveals specific structural patterns, as well as nodes contributing significantly to the most localized eigenvector
being crucial for the occurrence of the disease. 

\section*{Materials and Methods}
\subsection*{Data assimilation and network construction}
We study the PPI network of DM-II, where nodes are the proteins and edges denote the interactions between these proteins. 
After diligent and enormous efforts, we collect the protein interaction data from various literature and bioinformatic sources. 
To keep the authenticity of the data intact, we only take the proteins into account which are reviewed and cited. 
We consider two widely used bioinformatic databases namely Genbank from NCBI \cite{Genbank} and UNIPROT 
constituting data available from other resources like European Bioinformatic Institute, 
the Swiss Institute of Bioinformatics, and the Protein Information Resource \cite{UniProt}. 
To add more information, we take highly studied diabetes mellitus II cell lines 
whose protein expression data is known.  
Here, we use the protein expression data of EndoC$-$ $\beta$ H$1$ cells in the disease dataset \cite{gordon}.
Since there are a very few human cell lines available, we take the mouse and rat 
cell lines namely MIN and INS-1, proteins of which have already been proved to behave similarly in humans \cite{cell,waanders}. 
After collecting the proteins for both the datasets, their interacting partners are downloaded from STRING database \cite{STRINGdatabase}.
We take into account largest sub-networks from the normal as well as from the disease data and investigate their structural and spectral properties.

\subsection*{Structural measures}
There are several structural measures used 
enormously in the last fifteen years in order to understand and characterize the properties of a network \cite{Barabasi_2002}. 
Here we define only those which are useful for our analysis.
To start with, let us first define the interaction matrix or the adjacency matrix of the network as follows:
\begin{equation}
A_{\mathrm {ij}} = \begin{cases} 1 ~~\mbox{if } i \sim j \\
0 ~~ \mbox{otherwise} \end{cases}
\label{adj_wei}
\end{equation} 
The most basic structural parameter of a network is the degree of a node ($k_i$),
which is defined as the number of neighbors the node has ($k_i=\sum_j A_{ij}$). 
The degree distribution $p_({k})$, revealing the fraction of vertices with degree $k$, 
is known as the fingerprint of the network \cite{Barabasi_2002}. 
Another important parameter is the clustering coefficient (CC) of the node $(i)$,
which is defined as the ratio of the number of connections a particular node has and the possible number of 
connections that particular node can have. Clustering coefficient of a network can be written as 
\begin{equation}
\langle CC \rangle = \frac {1} {n} \sum_{i=1}^{n} CC_i
\label{eq_clique}
\end{equation} 
The average clustering coefficient of the network characterizes the overall 
tendency of nodes to form cluster or groups.
Further, the betweenness centrality of a node is defined as the fraction of shortest paths 
between node pairs that pass through the said node of interest \cite{Newman_2003}.
Another important parameter is the diameter of the network which measures the longest of the shortest paths 
between all the pair of nodes \cite {Barabasi_2002}. 

\subsection*{Spectral techniques}

Eigenvalues of adjacency matrix (Eq.~\ref{adj_wei}) denoted as $\lambda_1>\lambda_2> \lambda_3> \hdots > \lambda_N$
is referred as network spectra. Further, we use the inverse participation ratio (IPR) to analyze 
localization properties of the eigenvectors. The IPR of $k$th eigenvector $E^k$ with its $l$th
component being denoted by $E_l^k$, can be defined as 
\begin{equation}
I^k = \frac{ \sum_{l=1}^{N} [E_l^k]^4}{ (\sum_{l=1}^{N} [E_l^k]^2)^2}
\label{eq_IPR}
\end{equation}
which shows two 
limiting values : (i) a vector with identical components $E_l^k \equiv 1/\sqrt{N}$ has $I^k=1/N$, whereas (ii) 
a vector, with one component $E_1^k=1$ and the remainders being zero, has $I^k=1$. Thus, the IPR quantifies the 
reciprocal of the number of eigenvector components that contribute significantly.
An eigenvector whose components follow Porter-Thomas distribution yields $I^{k}$=$3/N$ \cite{Haake},
and those which are deviating from this value provide system dependent information \cite{sj_2012}.
We further calculate the average value of IPR, in order to measure an overall
localization of the network calculated as 
\begin{equation}
\langle IPR \rangle = \frac{ \sum_{k=1}^{N} I^k}{N}
\label{eq_IPRavg}
\end{equation}

\begin{table}[!ht]
\begin{center}
\caption{\bf{Detailed parameters for normal and disease sub-networks}}
\begin{tabular}{|c|c|c|c|c|c|c|c|c|c|c|}    \hline
Network & $N$ & $N_{C}$ & $\langle k \rangle$ & $D$  & $\langle CC \rangle$   & $N_{clus=1}$ & $\langle IPR \rangle$ &  $\lambda_{0}$ & $\lambda_{-1}$ \\ \hline
Normal-1 & 2083	& 11017	&  10   &  17	& 0.35	&  6.1$\%$	&   0.010 &  3.1$\%$ & 1.8$\%$ \\ \hline 
Normal-2 & 1705	& 9888	&  11   &  12 	& 0.33	&  6.2$\%$	&   0.008 &  3.5$\%$ & 1.8$\%$ \\ \hline 
Disease-1 & 656 & 3628	&  11	&  10	& 0.36	&  6.0$\%$	&   0.015 &  3.6$\%$ & 1.06$\%$ \\ \hline 
Disease-2 & 384 & 1882	&  10	&   9	& 0.46	&  7.8$\%$	&   0.032 &  7.2$\%$ & 4.9$\%$ \\ \hline 
\end{tabular}
\begin{flushleft}The columns represent the total number of proteins (nodes) in the network $N$ collected using
various databases (described in the Method section), number of connections in the sub-networks $N_{C}$, 
the average degree $\langle k \rangle$, diameter (D), average clustering coefficient $\langle CC \rangle$, 
the number of nodes having $CC=1$ in the whole network ($N_{clus=1}$), average IPR $\langle IPR \rangle$, 
and number of degenerate eigenvalues in the whole network $\lambda_{0}$ and $\lambda_{-1}$ for the sub-networks of both the normal and the disease state.
\end{flushleft}
\label{Table 1}
\end{center}
\end{table}

\section*{Results and Discussions}
\subsection*{Structural properties}
The normal dataset has 4613 nodes (proteins) and 26035 connections (interactions) among them, 
while the disease dataset comprises of 1100 nodes and 5578 connections 
exhibiting that the disease data consists of less 
number of proteins as well as connections than the normal one. 
One of the possible reasons behind this could be 
less availability of data for the disease state as already discussed 
under the data assimilation section. Another possible reason could be that
in the disease state, many pathways are silenced or levels of protein expressions are altered leading to
less number of proteins \cite{barabasi}.
The normal and the disease datasets lead to many connected clusters or sub-networks, 
whose sizes are summarized in the first column of Table.~\ref{Table 1}. 
The next column in Table.~\ref{Table 1} conveys that even 
with less number of nodes, the disease sub-networks possess almost same ($\langle k \rangle$) as the normal ones.
Further, the disease sub-networks show relatively lower diameter as compared to the normal one.
Since small diameter facilitates a fast communication \cite{Watts}, up-regulation and down-regulation of pathways may be one of the
reasons behind faster signaling leading to the disease state.
Further, we calculate the average clustering coefficient of all the sub-networks along with 
the total number of nodes having $CC$ equal to $one$ ($N_{clus=1}$) 
as tabulated in Table.~\ref{Table 1}.
The percentage of nodes having this property for the normal and all the disease sub-networks,
except $D-2$, are approximately same. For $D-2$, the number of nodes having $CC=1$ is slightly higher
as compared to other sub-networks indicating the presence of more clique structures in this sub-network.
Importance of cliques and relevance of this structure for the disease will become more clear in the section 
``preserved structure through clique formation" section.

\begin{figure}[!ht]
\centerline{\includegraphics[width=0.46\columnwidth]{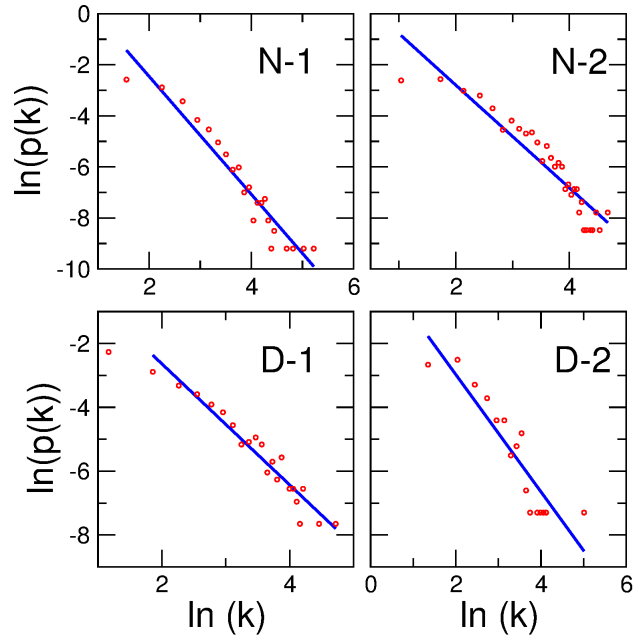}}
\caption{{\bf Degree distribution.} Degree distribution for the normal ($N-1$ and $N-2$) and disease ($D-1$ and $D-2$) datasets following the power law behavior of GOE statistics.}
\label{Figure 1}
\end{figure}

The degree distribution $p(k)$ of 
both the normal and the disease sub-networks follow power law (Figure.~\ref{Figure 1})
indicating that few nodes have very high degree.
Earlier studies on structural analysis of DM-II network also
emphasize on proteins having large number of interactions being functionally important \cite{Hongfong2008}. 
Since the present investigation uses a much larger number of proteins and interactions as
compared to any other work done on DM-II, our
datasets reveal new proteins that comes out to be top in the degree distribution. The bigger dataset used in 
our investigation arises due to the following two reasons, first, we have made
exhaustive literature search for almost all the authenticated databases to construct our dataset, and second, there is
revelation of new proteins and interactions since 2008 \cite{Hongfong2008} contributing to additional interactions
to the proteins in our dataset.
 
In the first disease sub-network, UBB and UBA-52 are the highest degree nodes, while  
for the second sub-network, UBI-15, INS, ALB and BBS-12 come out to be significant. 
We find that, these nodes are involved in insulin production and are known to be regulated in the disease
state leading to less production of insulin.
The detailed functional description of high degree nodes is deferred to the supplementary material.
Further, since the power law degree distribution is known to confer
robustness to the underlying network against random external perturbations as well as 
to instill complexity in the corresponding system \cite{Barabasi_2002}, 
the similar behavior exhibited by the disease and normal networks bring them into 
the same universality class of other complex biological systems \cite{Barabasi_correlation}.
What follows that, overall structural properties such as diameter, clustering coefficient,
average degree and degree distribution
of disease and normal sub-networks are similar indicating complex interactions and fast
dissemination of information in the pancreatic cell which is in the line of other complex systems.
The important differences between the disease and normal states are revealed when we analyze nodes forming complete sub-graphs
as well as those appearing in spectral analysis under the RMT framework.

\subsection*{Preserved structures through clique formation}

All the sub-networks have very high value of $\langle CC \rangle$ 
as depicted in Table.~\ref{Table 1}. This implicates the presence of high 
number of triangles or cliques of order three \cite{Watts}. 
Cliques indicate preserved interactions in the networks and are believed to be conserved during evolution \cite{clique_mutation}. 
Further, these structures are also considered to be the building blocks of a network for making
the underlying system more robust \cite{Alon} and stable \cite{skd}.
Therefore, we hope to bring out important information by analyzing these patterns in details. 
We focus on the nodes 
having $CC=1$ and which are common in both the networks.
It turns out that, there are 
34 nodes appearing common in the normal and disease datasets.
We perform extensive functional analysis of these nodes to study the differences
in both the states as these proteins enjoy special structural feature in the networks.

\subsection*{Biological significance of proteins involved in preserved structures}

The functional properties of these nodes reveal that out of 34 there are six proteins, namely 
ADAMTS9, G5A, PIK3C2A, VAPA, SPTBN2 and CTNNB2, 
which are responsible for regulating the levels of insulin in the blood leading to insulin resistance 
which is a primary symptom of the occurrence of type 2 diabetes in an individual \cite{relation_Insulin}.
Another set of proteins having $CC=1$ namely NR2C2, AGER, CACNA1D and CD59 are involved in the abnormal secretion 
and absorption of glucose which is another reason for DM-II. 
Some proteins viz. PTPRN, TGM2, SCARB2, VAPA, CBX3 and SPTBN2 are involved both in the 
insulin resistance and aberrant glucose metabolism \cite{relation_glucose_metabolism}.
Further, TAGLN and VDAC2 are two proteins which are found up-regulated in obesity and 
as obesity is strongly associated with DM-II, these proteins stand important 
for occurrence of the disease. SLC16A1 is associated with regulation of lipid metabolism 
and is also a novel protein for diabetes mellitus II \cite{relation_Obesity}.
Proteins OTUB1, UCHL1, NQO1 and PSMC5 have significant roles in 
ubiquitin-proteosome system which is regulated in DM-II. 
Next, expressions of SLC3A2 and IGHA1 are altered which are associated with excess urination
and thereby leading to the disease state as excessive urination is a major symptom of DM-II.
Moreover, VDAC2 and HPRT1 play a major role in house-keeping functions of the cell by energy regulation. 
Other proteins C6, C8A, NME3, DDC, FBLN1 and EKB are not directly related with the onset of the disease 
and perform house-keeping functions. Elaboration on above analysis is given in supplementary material. 
What follows that, out of $34$, $28$ proteins are directly
responsible for the occurrence of Diabetes mellitus II. 
The nodes having $CC$ equals to $one$ common in both the disease and normal 
have a significant role in the occurrence of the disease.
We further compare these structural properties to various other model networks namely random, 
small world network and configuration networks to examine the deviation of
the disease and normal networks from the random controls.

\begin{table}[!ht]
\begin{center}
\caption{\bf{The corresponding Erd\H{o}s-R\'{e}nyi model for all the normal and disease dataset}}
\begin{tabular}{|c|c|c|c|c|c|c|c|c|c|}
\hline 
Network & $\langle CC \rangle$ & $N_{clus=1}$ & $\lambda_{0}$ & $\lambda_{-1}$\\ 
\hline 
Normal-1 & 0.005 & 0 & 0 & 0 \\ 
\hline 
Normal-2 & 0.007 & 0 & 0 & 0 \\ 
\hline 
Disease-1 & 0.017 & 0 & 0 & 0 \\ 
\hline 
Disease-2 & 0.027 & 0 & 0 & 0 \\ 
\hline 
\end{tabular}
\begin{flushleft}The parameters show all the results as ensemble average of 10 realizations of Erd\H{o}s-R\'{e}nyi network with 
same $N$ and $\langle k \rangle$ as for real networks.
\end{flushleft}
\label{Table 2}
\end{center}
\end{table}

\subsection*{Comparison with various random control networks}
\subsubsection*{Erd\H{o}s-R\'{e}nyi random model}

Erd\H{o}s-R\'{e}nyi random network is the simplest model proposed about half a 
century back, to understand structural properties of real world systems \cite{ER}. 
We generate Erd\H{o}s-R\'{e}nyi network with the same number of nodes and average 
degree as the normal and the disease sub-networks.
To create an Erd\H{o}s-R\'{e}nyi network, we connect nodes with a fixed probability $p$, calculated as $\langle k \rangle/N$.
Since nodes are connected randomly with the probability $p$, without any special preference 
or additional constraints, this gives rise to the Poisson degree distribution. 
Properties of Erd\H{o}s-R\'{e}nyi random networks thus constructed, are summarized in Table.~\ref{Table 2}. 
Owing to the very much nature of construction of the Erd\H{o}s-R\'{e}nyi networks, we get
very small clustering coefficient ($CC \sim k/N$) which further
diminishes the probability of having any node with $CC=1$.
Additionally, there is no degeneracy found at $0$ or $1$ eigenvalues.

\begin{table}[!ht]
\begin{center}
\caption{\bf{Small world network}}
\begin{tabular}{|c|c|c|c|c|c|c|c|}
\hline
Network & $\langle CC \rangle$ & $N_{clus=1}$ & $\lambda_{0}$ & $\lambda_{-1}$\\
\hline
Normal-1 & 0.36 & 0  & 0 & 0 \\
\hline
Normal-2 & 0.36 & 0  & 0 & 0 \\
\hline
Disease-1 & 0.32 & 0 & 0 & 0 \\
\hline
Disease-2 & 0.44 & 0 & 0 & 0 \\
\hline
\end{tabular}
\begin{flushleft}Various parameters enlisted for small-world 
networks corresponding to the normal and disease sub-networks taking ensemble average of 10 realizations .
\end{flushleft}
\label{Table 3}
\end{center}
\end{table}

\subsubsection*{Small-world network}
Since all the disease networks have significantly high value of $\langle CC \rangle$ indicating
hidden importance of network interactions contributing to this property,  
we attempt to model this and investigate using
corresponding small world networks (Table.~\ref{Table 3}).
The small world networks are generated using Watts-strogatz algorithm \cite{Watts}. Starting from a regular lattice,
connections are rewired with the probability $p$. The rewiring probability
is chosen in such a manner that it leads to the desired
clustering coefficient matching with
the corresponding real networks. It turns out that while for the model
network contribution to the CC comes from almost all the nodes, for
the real world networks only $75\%-80\%$ nodes contribute to the clustering coefficient
of the network and rest of the nodes do not have any interacting neighbors. i.e. they
form star like structure. Additionally, real networks have significant 
number of nodes having $CC=1$, the property which again is not found
in the corresponding model network. 
As elaborated in the previous section ``preserved structure through clique formation", the nodes 
with $CC=1$ have special role in the occurrence of the disease, making the deviation of $CC=1$ properties of the nodes
from the model networks more crucial.

\begin{table}[!ht]
\begin{center}
\caption{\bf{The configuration model for corresponding real world dataset}}
\begin{tabular}{|c|c|c|c|c|c|c|c|}
\hline 
Network  & $\langle CC \rangle$ & $N_{clus=1}$ & $\lambda_{0}$ & $\lambda_{-1}$\\ 
\hline 
Normal-1 & 0.023  & 7 & 1.5$\%$ & 0 \\ 
\hline 
Normal-2 & 0.027  & 4 & 1.7$\%$ & 0 \\ 
\hline 
Disease-1 & 0.083 & 12 & 4.5$\%$ & 0 \\ 
\hline 
Disease-2 & 0.123 & 7 & 2.1$\%$ & 0 \\ 
\hline 
\end{tabular}
\begin{flushleft}Ensemble average of 10 realizations for parameters of configuration model corresponding to the normal and diabetes networks listed for various properties. 
\end{flushleft}
\label{Table 4} 
\end{center}
\end{table}

\subsubsection*{Configuration model}
Erd\H{o}s-R\'{e}nyi random and small world networks being the 
simplest models for capturing various properties of real systems fail to model
one of the very important aspects of many real-world networks.
The degree distribution of Erd\H{o}s-R\'{e}nyi random networks follow Poisson distribution
whereas most of the real world networks \cite{Barabasi_2002}, including those investigated here,
follow power law behavior. 
This power law behavior can be modeled using preferential attachment rule \cite{barabasi_1999}, 
but configuration model additionally preserves the exact degree sequence of a network \cite{conf_model} and hence
we make comparison with this model which produces an exact random replication of the network considered here.
The configuration model generates a random network with a given degree sequence of an array of size 
$m=\frac{1}{2}{\sum_{i=1}^Nk_i}$ which have random connections among different elements.
We generate 10 such realizations for a given degree sequence.

Properties of the configuration model corresponding to the
normal and diabetes sub-networks are enlisted in Table.~\ref{Table 4}. The clustering coefficient, inverse participation 
ratio, of all the configuration model networks are all same as for the Erd\H{o}s-R\'{e}nyi random networks
as expected from the configuration model algorithm, which randomly connects pair of nodes
by following only one constraint that is preservation of the degree of all the nodes.
For configuration model, there is an additional property which comes out to be different
than other model networks that is about degenerate eigenvalues. 
This occurrence of $zero$ degeneracy is not surprising \cite{Farkas} as power law degree distribution
is known to add into occurrence of high degeneracy at $zero$ \cite{jost}. 
Further, we note that all these model networks, while acting under one or the other constraint,
are basically built up on random interactions, i.e., the pairs of nodes that should interact are
chosen randomly out of many possible configurations.
The feature which real-world systems are supposed to lack and in turn being captured in various spectral properties.
For instance, none of the random controls are able to reproduce high degeneracy at $zero$. 
In addition, there is degeneracy at $one$ as well, which too is not captured by
any of the model networks widely investigated in the network literature reflecting 
richer interaction patterns possessed by these real systems.
In the next sections, we will delve into understanding the spectral features of the four sub-networks.

\begin{figure}[!ht]
\centerline{\includegraphics[width=0.46\columnwidth]{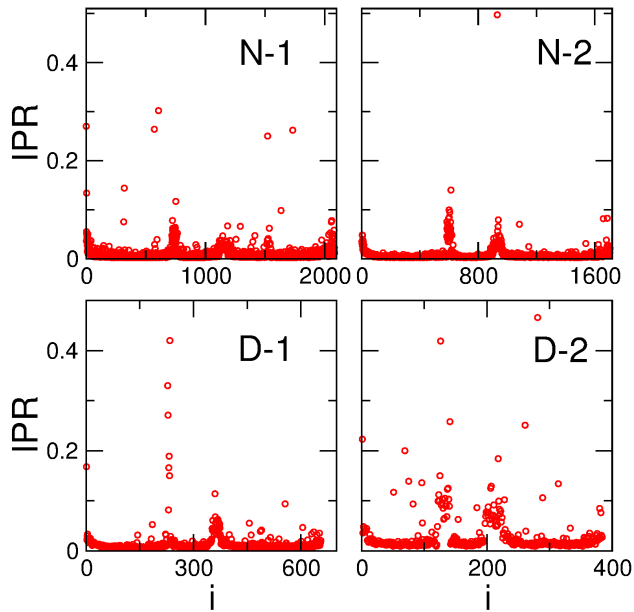}}
\caption{{\bf Eigenvector localization for both the normal and disease sub-networks.} The figure shows IPR of both normal ($N-1$ and $N-2$) and disease ($D-1$ and $D-2$) network, clearly
reflecting three regions (i) degenerate part in the middle, (ii) a large
non-degenerate part which follow GOE statistics of RMT and (iii)non-degenerate
part at both the end and near to the zero eigenvalues which deviate from RMT.}
\label{Figure 2}
\end{figure}

\subsection*{Localization behavior of spectra}

The localization properties of eigenstates have been shown to provide clue to the important nodes of the network \cite{SJ_epl2009}.
We calculate the eigenvector localization using IPR (Eq.~\ref{eq_IPR}), and based on the
localization behavior, we can divide the individual spectrum into two different components. 
First component which follows RMT predictions of Porter-Thomas distribution of GOE statistics \cite{PT},
and another one which deviates from this universality and indicates localization (Figure.~\ref{Figure 2}).
The very distinct interpretation of this behavior of universal and non-universal (deviation from universality) components
is that the underlying system has random interactions leading to universal part of the spectra, as well as non-random, 
pattern specific interactions contributing to the part of the spectra deviating from the universality.
Further, the average IPR calculated using Eq.~\ref{eq_IPRavg}
for the normal and disease sub-networks reveal that the sub-networks of normal are less 
localized than the sub-networks of disease 
networks (Table.~\ref{Table 1}) indicating that the normal sub-networks are more random than the disease sub-networks.
The fact is that the disease state has less number of connections than that of the normal one
suggests that there are some interactions which are hampered or silenced perhaps
due to mutation leading to the disease state. 
This interpretation combined with the analysis that $\langle IPR \rangle$ of the disease is higher than the normal,
indicates that these hampered pathways may be corresponding to
or should be treated as random pathways whose removal result in the diminished randomness in the disease state.
This result can be considered very important as {\it randomness} has already been emphasized
as an essential ingredient for the robustness of a system \cite{SJ_epl2009} and
lack of {\it sufficient randomness} might lead to the disease state.  

\begin{table}[!ht]
\begin{center}
\caption{\bf{TCNs of Disease-1 sub-network}}

    \begin{tabular}{ | p{0.6cm} | p{6.0cm}| p{2cm}| p{2cm}| p{2cm}|} \hline
$ E^k $ &  TCN &	k & CC  & BC \\ \hline
234&	PSMD1, PSMD11, PSMB8 &	4, 4, 4 & 1, 1, 1 & 0, 0, 0
	\\ \hline
228&	PFKP, PFKL, PSMB8, H3.3B & 5, 5, 4, 8 & 0.9, 0.9, 1, 0.75 & 0, 0, 0, 0.002
		\\ \hline
229&	COL1A2, COL3A1, PSMB8, H3.F3B & 6, 6, 4, 8 & 0.5, 0.53, 1, 0.75 & 0.002, 0.002, 0, 0.002  
	\\ \hline
232&	PSMB8, PSME1, PSMD11, H3F3B, H3.3B, PSMD1 & 4, 4, 4, 8, 8, 4 & 1, 1, 1, 0.75, 0.75, 1 & 0, 0, 0, 0.002, 0.002, 0
	\\ \hline
\end{tabular}
\begin{flushleft}Top localized eigenvectors ($E^k$) for the disease $D-1$ dataset 
representing the index of the localized eigenvector followed 
by their TCN's in the next column and network parameters namely degree, 
clustering coefficient and betweenness centrality.
\end{flushleft}
\label{Table 5}
\end{center}
\end{table}

\begin{table}[!ht]
\begin{center}
\caption{\bf{TCNs of Disease-1 sub-network}}
    \begin{tabular}{ | p{0.6cm} | p{6.0cm}| p{2cm}| p{2cm}| p{2cm}|} \hline
$ E^k $ &  TCN &	k & CC  & BC \\ \hline
281&	AVPR2, MT-ND5, MT-ND4 &	2, 13, 13 & 1, 0.846, 0.846 & 0, 0, 0.003
	\\ \hline
126&	MT-ND1, MT-CO3, MT-ND6 & 	12, 1, 12 & 0.985, 0.985, 0.984 & 0, 0, 0
	\\ \hline
141&	SEL1L, SELS, C1S, HLA-DQA1 & 1, 2, 9, 6 & 0, 0, 0.5, 0.46 & 0, 0.005, 0.004, 0
	\\ \hline
261&	YWHAE, CALM1, CALM3, CALM2 & 1, 6, 5, 5 & 0, 0.53, 0.80, 0.80 & 0, 0.009, 0.003, 0.003
	\\ \hline
\end{tabular}
\begin{flushleft}Top most localized eigenvectors ($E^k$) for the disease $D-2$ dataset, their top contributing
proteins and network parameters namely degree, clustering coefficient and betweenness centrality.
\end{flushleft}
\label{Table 6}
\end{center}
\end{table}

As discussed, eigenvector localization technique offers the platform to distinguish
random and non-random part of the spectra. The part deviating 
from RMT can be explored further to get insight
into the important interactions revealed through the top contributing nodes (TCNs).  
In order to do that, we investigate properties of TCNs of both 
the sub-networks $D-1$ and $D-2$ as described in Table.~\ref{Table 5} and Table.~\ref{Table 6} respectively.
There comes out to be 31 TCN's in the disease sub-networks out of which 24 are unique.
We analyze the structural properties namely, degree, 
clustering coefficient and betweenness 
centrality of these TCN's in both the sub-networks. 
The top contributing proteins 
lie in the low degree regime demonstrating that they do not take part into many pathways. 
Additionally, the betweenness centrality of all the TCNs is nearly zero, indicating a very poor
connectivity of these nodes with the rest of the nodes in their individual sub-network.
In coalition, these both rule out a trivial importance of the nodes in terms of their interactions or connectivity,
and hence we shift our focus to the biological significance of these 
TCNs and interaction patterns they are forming in the network.

\subsection*{Functional importance and interaction pattern of top contributing nodes}

\begin{figure}[!ht]
\centerline{\includegraphics[width=0.6\columnwidth]{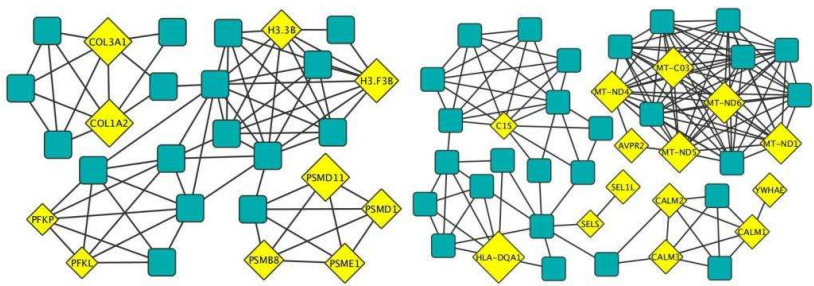}}
\caption{{\bf Interaction patterns of TCNs in disease datasets.} 
Left panel depicts the local structure of all TCNs in the $D-1$ sub-network.
Right panel has the interaction pattern of TCNs in $D-2$ sub-network where Diamonds resemble the top contributing nodes and rectangular boxes are the first interacting partners of these TCN's.}
\label{Figure 3}
\end{figure}

These TCNs when looked for their functional properties are found to be significantly important in DM-II. 
Among all the TCNs, SEL1L, SELS, PFKL and PFKP are associated
with the glucose stimulated insulin secretion. In disease state SEL1L and SELS proteins 
are found to be under-expressed \cite{Adam_2011} whereas PFKL and PFKP over-express \cite{Elson_1994}. 
In addition to the above proteins, C1S, YWHAE, CALM1, CALM2, CALM3 and MT-CO3 are responsible 
for obesity in turn regulating levels of insulin leading to insulin resistance in the body and 
thus contributing to the occurrence the disease \cite{Jinhui_2007,Adeyemo_2009,Ozcan_2013}.
The proteins PSMD1, PSMD11, PSME1 and PSMB8 form a complex in the proteosome 
accounting for ATP dependent protein degradation and in disease state are found to be down regulated or absent
which is directly related to obesity and DM-II \cite{Aldrin_2013,Sakamoto_2009}. 
Further, MT-ND5, MT-ND4, MT-ND1 and MT-ND6 collectively are the components in mitochondria 
involved in insulin signaling and resistance as well as alteration in the cell leading to DM-II \cite{Michal_2007}.
Additionally, increased levels of AVPR2 leads to glucose induced water loss in the body, hence
leading to the disease state \cite{Ineke_2013}. Protein HLA-DQA1 plays a central role 
in immune system and over expression of this protein 
increases the susceptibility to DM-II \cite{Ze-Jun_2013}.  
Furthermore, there are some TCNs localized but are not directly involved with DM-II namely COL1A, COL3A1, H3.3B and H3F3B. 
These proteins are found to play an important role in complications related to heart, lung and kidneys in turn 
leading to the symptoms of DM-II like regulated levels of glucose absorption in blood and 
urinary complications \cite{Lunteren_2013,Bhanudas_2010}.
A detailed functional information of all these proteins are discussed in the supplementary material.

Next, we analyze the structural patterns of the TCNs in both the disease sub-networks.
$D-1$ sub-network reveals the formation of cliques of order three for all the top contributing nodes (Figure.~\ref{Figure 3}). 
This depicts that they are all involved in the same kind of pathways which is confirmed
by studying the functional properties of these proteins briefly discussed above in this section.
Further, the TCNs of $D-2$ sub-network too comprise of complete subgraphs (Figure.~\ref{Figure 3}). 
The functional properties discussed above demonstrates that these TCNs form 
complexes as they perform same kind of functions
e.g. proteosome complex (PSMD1, PSMD11, PSME1 and PSMB8), mitochondrial components (MT-NDs), 
collagen family (COL3A1 and COL1A) and histone complexes (H3.3B and H3F3B), details of which are explained in supplementary material.
Thus, the interaction patterns of both the networks of disease dataset reveal that
there are nodes pertaining to clique formations which indicate a robust  
and stable system as discussed in the section on ``preserved structure through cliques". 

\begin{figure}[!ht]
\centerline{\includegraphics[width=0.6\columnwidth]{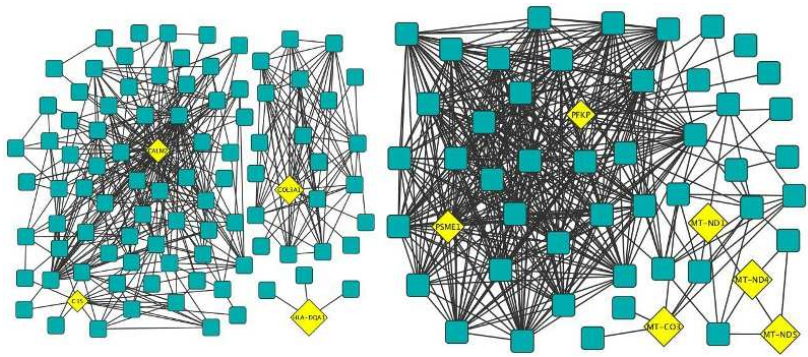}}
\caption{{\bf Interaction patterns of disease TCNs in normal datasets.}
The local structural patterns for the disease TCNs in the normal network. 
The left and right panels represent $N-1$ and $N-2$ sub-networks respectively resembling  
Diamonds as the top contributing nodes and rectangular boxes as the first interacting partners of TCN's.
The same TCNs in normal sub-networks possess more interactions than the disease sub-networks.}
\label{Figure 4}
\end{figure}

Thereafter, we compare the interaction patterns of TCNs in the disease state 
with their interaction patterns in the normal.
Out of 24 TCNs in disease state, 10 are also found in the normal dataset. 
The interaction pattern of the disease TCNs in normal sub-networks possess clique 
like structures. The TCNs in normal dataset have comparatively high degree than 
in the disease state indicating more interactions. This may be due to silencing
of various pathways (e.g. insulin signaling, glucose metabolism etc) from normal to the disease state
as discussed earlier.
The structural patterns indicate that even after less interactions in the disease state, 
the cliques of order three are not destroyed reflecting that the minimum criteria 
of maintaining the functionality of the networks are preserved.
This may be due to addition of new proteins in the diseased state or interactions among themselves.  
In addition, the interacting patterns of these TCNs illustrate the phenomena of gene duplication 
i.e. a pair of node having exactly same interacting partners \cite{Ohno}. 
Here, out of the 10 TCNs common in both the states, 
two pairs of nodes namely C1S-CALM2 and PSME1-PFKP in the normal dataset 
possess the phenomena of gene duplication which is lost in the disease state (Figure.~\ref{Figure 4}).
Gene duplication has structural and functional significance as fluctuations or destruction in the interactions 
can either cause imbalance of genetic material or lead to the generation of new gene products
resulting in diseases \cite{JR}.  
In TCNs, the proteins possessing gene duplication behavior
affects the major pathways namely insulin resistance and pathways promoting obesity in turn resulting in DM-II.
Thus, proteins revealed through the localization properties of spectra have interaction patterns 
which may be important as targeting sites for drug development of DM-II.

\subsection*{Conclusions}

We analyze the pancreatic cells for the normal and disease states to get a deep overview of DM-II under the combined
framework of network theory and random matrix theory. 
The structural properties namely average degree, diameter, clustering coefficient and degree distribution 
of both the networks follow almost similar predictions 
as for the other biological networks \cite{Barabasi_correlation}.
The average degree of all the normal and disease sub-networks is nearly same indicating that 
the disease state has some altered interactions and some new connections.
The diameter of disease sub-networks is less than the normal which indicates that a reaction may get completed in few steps 
in turn increasing the rate of reaction and faster communication in processing a biological function.
The degree distribution of all the normal and the disease sub-networks follow
the power law behavior,
the property which is known to make a system robust \cite{Barabasi_correlation}.
Interestingly, despite so many alterations in the interactions, 
disease networks still exhibit this property possessed by the normal networks indicating
robustness and complex nature of interactions in the disease state.
This could be one of the possible reasons of failure or poor response of drugs
for DM-II \cite{kitano}.

The interesting observation comes out from the nodes having $CC$ equal to $one$ indicating
existence of clique structure in the network and since
cliques are considered as preserved structures during the evolution    
making the system more robust and stable, we investigate biological functions of these nodes.
It turns out that the nodes having $CC$ equals to $one$ play an important role in the occurrence of the disease. They are involved in 
important pathways like insulin resistance, glucose metabolism and obesity which are directly related to type II diabetes.

Further, we compare properties of normal and disease sub-networks with those of
various model networks to get a deeper understanding of various characteristic properties of networks under consideration.
While corresponding model networks capture one or many structural properties of real world networks,
the high degeneracy of eigenvalues found for the real networks are not observed for any of the 
model networks. Since, model networks are built using some algorithm, which connects
pair of nodes using some probability distribution, deviation of this property of real world networks
from that of model networks reflect that there are non-random interactions in the real world system 
which in turn leads to the formation of patterns, hence affecting the spectra of the network,
leading to specific deviation of its properties from the corresponding random controls.

Furthermore, the spectral analysis of both the networks, by calculating
the localization properties of eigenvector through IPR, 
reveals that the the disease sub-networks are less random than the normal one. 
Randomness in network connections or interactions has already been emphasized for the proper functioning of various complex systems
\cite{strogatz_2001,Cohen,treves,ohtsuki,sinha}. 
The revelation that the disease network is less random than the normal one, 
detected through the RMT techniques, is an important step to complex systems research at the fundamental level. 
It indicates that sufficient randomness is an essential ingredient for proper functioning of the system
leading to a normal state, lack of which causes the disease.
The TCNs of the most localized eigenvectors lie in the low degree regime and have
almost zero betweenness centrality, exhibiting no trivial structural importance. 
But they form important structural patterns as out of $24$ TCNs, $21$ have been found to 
form cliques with other proteins. 
Since, clique structures have been emphasized to be important for robustness of the underlying system \cite{Alon},
the proteins forming cliques here indicate that as
random perturbations in the network leaves the system stable. 
This may also lead to ineffective drugs and treatments to the patients 
as also predicted from the power law behavior discussed earlier in this section.
The functional examination of these TCNs demonstrates that out of $24$, $20$ proteins have 
biological significance in the occurrence of the disease by involving in crucial pathways 
related to DM-II such as insulin signaling and resistance. These structural and functional implications
reflect that the localization property reveals important nodes of the network. 

We investigate interactions of TCN in normal dataset as well.
It is found that 10 TCNs from the disease dataset are common in the normal dataset. 
The connections among common TCNs form the same conserved structure i.e.
clique. In addition, there are two pairs of TCNs in the normal dataset 
which follow the phenomena of gene duplication which is lost in the disease network.
It is already known that the alterations or destructions in the interactions having gene duplication leads to disease state \cite{JR}.
On one hand, the revelation that the number of interactions of TCNs in normal dataset is more than that those
observed in the disease indicates silencing of pathways in the disease state.
On the other hand, comparison of structural patterns of TCNs in the normal and disease states
reveals that even after removal of various connections probing to gene duplication either
by forming new connections with new proteins or within themselves in the disease condition,
the underlying system subsists.
Thus, these patterns may play an important role in depicting novel drug targets taking them as site of action. 
It would be noteworthy to mention here that though the disease-implicating proteins revealed 
through our analysis belong to certain parts of the interaction network, 
they are detected through spectral analysis of the whole matrix and takes into account all 
interactions providing a holistic approach.

To conclude, we provide a platform to detect important proteins revealed through structural patterns
in DM-II using network approach and random matrix theory.
The analysis suggests that instead of targeting individual proteins, 
a group of proteins forming particular interaction patterns
should be taken into consideration for drug development.
This approach, though based on sophisticated mathematical techniques, is time-efficient 
and cost-effective and paves a way to look into the diseases
from a different perspective by taking the whole system into consideration. 
For the first time RMT is applied to analyze the DM-II and our work demonstrates that 
it can be modeled using appropriate ensemble of RMT. 
Being the first step, the approach holds the potential to provide a new dimension to disease research. 
RMT is a very well developed branch of physics and continues to witness emerging techniques and concepts \cite{rev_rmt1,goh1}.  
Appreciating the applicability of this technique in uncovering crucial information about DM-II, 
other well developed tools of RMT can be used to gain more insight into the complexity of DM-II and decipher important entities and their 
structural patterns significant in other diseases. 
Moreover, our work on one hand presents a new tool to identify proteins responsible for 
occurrence of the disease and predicts interaction patterns for drug targets, 
while on the other hand provides insight into complexity of the disease at the rudimentary level.
The analysis exemplifying the structural and functional significance of these proteins 
indicates that rather than individual units, the whole interaction pattern can be 
treated as a target for drug development providing a new direction to treatment of the disease.
This approach can also be extended in composition of novel drugs, 
conceptualization of single drug therapy for multiple diseases \cite{Goh,ma} and devising personalized therapy \cite{personalized_therapy} 
being especially beneficial for low-income countries.

\section*{Acknowledgments}
SJ is grateful to Department of Science and Technology, Government of India
and Council of Scientific and Industrial Research, India project grants 
SR/FTP/PS-067/2011 and 25(0205)/12/EMR-II for financial support.
AR is thankful to the Complex Systems Lab members Camellia Sarkar and Sanjiv K Dwivedi
for helping with the plots and useful discussions.

\subsection*{Author contributions} 
SJ conceived and supervised the project. AR collected the data and AKP searched functional properties.
SJ and AR analyzed the data and wrote the manuscript.

\subsection*{Competing financial interests}
The authors declare no competing financial interests.

\section*{Supporting Information Legends}
{\bf Supporting information file for the article ``Interaction patterns in diabetes mellitus II network: An RMT relation''}\\
\begin{description}
\item{{\bf Figure S1. Degree-CC Correlation.} The degree-CC correlation for the normal (N-1 and N-2) and disease (D-1 and D-2) networks showing negetive correlation and following power law.}
\item{{\bf Figure S2. Nodes having CC-1.} Degree distribution for the normal (N-1 and N-2) and disease (D-1 and D-2) datasets for nodes having CC=1.}
\item{{\bf Table S1. Biological functions.} Functional description of proteins having CC=1.}
\item {The supplementary file also accompanies functional properties of other common proteins having CC=1, 
the top degree node in disease and normal state followed by top contributing nodes 
of disease networks achieved through the localization properties in detail. }
\end{description}

\cleardoublepage
\centerline{\large \bf Supporting Information} 
\vspace{1cm}

\section*{Relation between diabetes mellitus type II and obesity} 

Obesity and T2d have a complex but direct relationsheep. Obesity stands out as a primary risk factor for T2D. According to its participation in T2D it can be assumed as a precursor for it, followed by insuline leading T2D 
\cite{Fagot_1998,Mitropoulos_1992,Frayn_1996}.

\section*{Degree clustering correlation}

Furthermore, the degree and CC of all the sub-networks of the normal as well as disease 
are negatively correlated (Figure.\ref{Figure S1}) 
as found for other biological systems. \cite{Zimmer}.
he degree and CC of the normal and the disease networks
are negatively correlated (Figure.\ref{Figure S1}), 
as found in the case of other biological systems. 
This negative correlation indicates that the low degree nodes tend to form modules 
which are connected through the high degree nodes for proper functioning of a system \cite{Barabasi_correlation1}. 
\begin{suppfigure}[ht]
\centerline{\includegraphics[width=0.4\columnwidth]{Figure_S1A.eps}}
\centerline{\includegraphics[width=0.4\columnwidth]{Figure_S1B.eps}}
\caption{{\bf Degree-CC Correlation.} The degree-CC correlation for the normal (N-1 and N-2) and disease (D-1 and D-2) networks showing negetive correlation and following power law.}
\label{Figure S1}
\end{suppfigure}

\section*{Degree distribution of nodes having CC=1}

Out of 72 proteins, the least degree a protein can have is two
as all these proteins have $CC=1$ indicating that they are
part of atleast a lowest order clique of size three.
There are 42 proteins have degree two. What follows that the lower degree proteins lying towards periphery
should correspond to end product. Since all these nodes having $CC=1$ have zero to close
to zero betweenness centrality confirm that this indeed is the case of the these nodes
and suggest that they are not involved in many pathways. Although there
are many proteins which correspond to end products lying
on the periphery of the network having degree one (Fig.~\ref{Figure S2}),
we concentrate
here on those proteins which in addition to have less degree
has property of having $CC=1$.

\begin{suppfigure}[ht]
\centerline{\includegraphics[width=0.4\columnwidth]{Figure_S2.eps}}
\caption{{\bf Nodes having CC-1.} Degree distribution for the normal (N-1 and N-2) and disease (D-1 and D-2) datasets for nodes having CC=1.}
\label{Figure S2}
\end{suppfigure}




\section*{Functional properties of common proteins having CC=1}

{\bf CTNNB1(Catenin(cadherin-associated protein), beta 1, 88kDa)-} It is a part of WNT signaling pathway which plays important role in several processes like adipogenesis, fat production, strong activator of mitochondrial biogenesis, also leads to increased  production of reactive oxygen species(ROS). This ROS-induced damage is significant because it can cause the development of acute hepatic insulin resistance, or injury induced insulin resistance. In diseased cells single nucleotide polymorphisms in β-cat/TCF lowers β-catenin levels which leads to the alteration or mutation of WNT signaling pathway make cells more susceptible to the development of T2D 
\cite{Zhai_2011,Agostino_2012,Logan_2004,Welters_2008,Yoon_2010,Grant_2006}.

{\bf Summary-} This protein is involved in WNT signaling pathway, which has important role in increased production of reactive oxygen species(ROS) which leads to insulin resistance (main cause of T2D). 
\\
{\bf TGM2(Transglutaminase 2)-} It is a multi-functional enzyme which catalyzes transamidation reactions or acts as a G-protein(guanosine nucleotide-binding protein) in intracellular signaling in normal cells. In disease cells lacking TGM2 makes them glucose intolerant and show impairment of insulin secretion, suggesting an important physiological role for TG2 in the pancreatic β cell \cite{Porzio_2007}.

{\bf Summary-} This acts as a G-protein in intracellular signaling in normal cells. Mutation in this leads to abnormal insulin secretion and glucose intolerance.
\\
{\bf SCARB2(Scavenger Receptor Class B, Member 2/ Lysosome membrane protein 2)-} It is a class B scavenger receptor normally functions as membrane protein whose expression is prevalent in vascular lesions. Raised levels of glucose, insulin resistance, low HDL cholesterol, increased levels of free fatty acid (FFA) all result in increased expression of CD36, thereby contributing to T2D. Adipocytokines such as tumor necrosis factor-alpha (TNF-α), C-reactive protein (CRP), adiponectin, leptin, resistin along with peroxisome proliferator activated receptor-γ (PPAR-γ) are important mediators in glucose homeostasis in association with CD36 and can be used as markers for T2D and atherosclerosis. Several of these gene variants have shown association with lipid metabolism, T2DM and related complications \cite{Gautam_2011}.

{\bf Summary-} It is a class B scavenger receptor, its increased expression leads to Raised levels of glucose, low HDL cholesterol, increased levels of free fatty acid (FFA), and insulin resistance.
\\
{\bf TAGLN(Transgelin)–} Expression level of this protein is up-regulated during obesity and leading T2D. A functional role of this protein is unclear. Two transcript variants encoding the same protein have been found for this gene \cite{Lee_2005,Entrez1}.

{\bf Summary-} Expression level of this protein is up-regulated during obesity and leading T2D.
\\
{\bf SPTBN2(Spectrin, Beta, Non-Erythrocytic 2 )-} It is a component of Cell membrane-cytoskeleton and is composed of two alpha and two beta spectrin subunits. Also regulates the glutamate signaling pathway and insulin signaling. In diseased cells due to increased protein glycosylation it is responsible for membrane abnormalities and develop insulin resistance causing T2D 
\cite{Deborah_1997,ROBERT_1991}.

{\bf Summary-} This protein regulates the glutamate signaling pathway and insulin signaling. In diseased cells due to increased protein glycosylation it is responsible for membrane abnormalities and development of insulin resistance causing T2D.
\\
{\bf VAPA(Vesicle-associated membrane protein-associated protein A)-} It is present in the plasma membrane and intracellular vesicles. It is associated with the cytoskeleton, vesicle trafficking, membrane fusion, protein complex assembly and cell motility, and also acts as receptor in Calcium signaling toolkit in normal cells. In diseased cells its expression got decreased with -1.35 folds which directly affects cell adhesion and vascular complications in T2D. any alteration in Ca2+ signaling adversely effects on cell survival and insulin secretion  leading T2D \cite{Kaviarasan_2009,Anubha_2012}.

{\bf Summary-} This protein playes important role in insulin secretion as receptor in Calcium signaling toolkit, and alteration in expression affects insulin secretion.  
\\
{\bf PTPRN(Receptor-type tyrosine-protein phosphatase-like N)-} It is involved in Ca2+ -regulated pathway as receptor in INS-1 cells which relates with glucose-induced exocytosis of insulin secretory granule (ISGs) in their biogenesis. Also regulate glucose stimulated β-cell proliferation. Destruction and dysfunction of the insulin-producing beta cells in the pancreas directly leads to T2D. And Deletion of PTPRN resulted in mild glucose intolerance and decreased glucose-responsive insulin secretion 
\cite{Kampf_2004,Warford_2004,Torii_2009,Cecilia_2012,Saeki_2002}.

{\bf Summary-} In ca2+ regulated pathway and glucose stimuilated B-cell proliferation, dysfunctioning or deletion makes cells glucose intolerance and decreased glucose-responsive insulin secretion and leads to T2D.
\\
{\bf PYCR1(Pyrroline-5-Carboxylate Reductase 1)-} It catalyzes the NAD(P)H-dependent conversion of pyrroline-5-carboxylate to proline in normal cells. its upregulation in high glucose conditions leads to the development of various diabetic complications including cardiovascular diseases, dibetic nephropathy, and retinopathy through upregulating cytokines such as TNF-a, IL-1B, IL-6, and MCP-1 in monocytes, by the IFN-a pathway which leads to T2D 
\cite{Brownlee_2001,Brownlee_2005,Giaccari_2009,Sheetz_2002,Dasu_2007,Devaraj_2005,Guha_2000,Jain_2007,Miao_2004,Shanmugam_2004,Shanmugam_2003}.

{\bf Summary-} This protein catalyzes conversation of proline in normal cells, and upregulation in diseased cells leads to alteration of IFN-a pathway and ultimately to T2D.
\\
{\bf OTUB1(Ubiquitin thioesterase OTUB1/otubain-1)-} It interacts with another protien ubiquitin protease and an E3 ubiquitin ligase that inhibits cytokine gene transcription in the immune system. It participates in specific Ubiquitin- Proteasome pathway /complex, which has a major role in targeting cellular proteins for ATP dependent degradation by the 26S. Any alteration or change in this pathway makes cell more susceptible to T2D \cite{Entrez2,Safia_2011,Rubinsztein_2006}.

{\bf Summary-} This protein interacts with ubiquitin protease and thus participate in Ubiquitin- Proteasome pathway, which playes major role in targeting cellular proteins for ATP dependent degradation by the 26S makes cell more susceptible to T2D.
\\
{\bf SLC3A2(4F2 cell-surface antigen heavy chain)-} In normal cells mainly functions for  renal handling of citrat, oxidative metabolism, codes for 4F2 heavy chain (activator of dibasic and neutral amino acid transporters) and also involved in the transport of the bulky branched-chain amino acids  which are very important in regulation of blood pressure and urine excretion and water re-absorption. 4F2hc can interact with GLUT1; such interaction promotes the stabilization of GLUT1 and contribute to the regulation of glucose metabolism. In diseased cells its expression gets upregulated and responsible for excessive urine excretion and alteration in SKG1 signaling pathway leading T2D \cite{Pajor_1999,Ho_2007,Bergeron_2011,Mastroberardino_1998,Meinild_2000,Diana_2013,Liqi_2014,Ohno_2011}.

{\bf Summary-} This protein mainly functions for renal handling, its increased expression leads to excessive urine excretion. 
\\
{\bf VDAC2(Voltage-dependent anion-selective channel protein 2 )-} It is a Part of voltage dependent anion channel (VDAC) which playes important role in cellular energy regulation, mitochondrial cell apoptpsis  in normal cells. And due to its functional importance it is directly associated with T2D and helps us in understanding of pathological conditions such as obesity and T2D 
\cite{Turko_2003,Mostyn_2004,McCabe_1994,Shoshan_2006,Lemasters_2006}.

{\bf Summary-} This protein is a part of voltage dependent anion channel (VDAC) which playes major role in cellular energy regulation, mitochondrial cell apoptpsis, and any alteration in functioning if this channel leads to  obesity and T2D.
\\
{\bf UCHL1(Ubiquitin Carboxyl-Terminal Esterase L1 (Ubiquitin Thiolesterase))-} It is  a thiol protease that hydrolyzes a peptide bond at the C-terminal glycine of ubiquitin, specifically expressed in the neurons and in cells of the diffuse neuroendocrine system in normal cells. It is an important component of the ubiquitin-proteasome system. It increases the available pool of ubiquitin to be tagged onto proteins destined to be degraded by the proteasome. In diseased cells Loss of Uchl1 increases the susceptibility of pancreatic beta-cells to programmed cell death, indicating that this protein plays a protective role in neuroendocrine cells and illustrating a link between T2D and neurodegenerative diseases. This gene is specifically expressed in the neurons and in cells of the diffuse neuroendocrine system. Mutation in this gene leads to T2D \cite{Osaka_2003,Entrez3,Chu_2012}.

{\bf Summary-} This protein is a Component of the ubiquitin-proteasome system, In diseased cells it increases the susceptibility of pancreatic beta-cells to programmed cell death, mutations in this gene leads to T2D. 

\begin{supptable}
\begin{center}
\begin{tabular}{| p{2.0cm} | p{10.0cm}|} \hline
 Protein  & Function \\ \hline
ADAMTS9  &  pancreatic beta cell dysfunction and decrease in insulin sensitivity.	\\ \hline
NR2C2	& down-regulation leading to hypoglycaemia which is directly related to TYD	 \\ \hline
C6	& gluconeogenesisleads to deficiency in insulin sensitivity 	\\ \hline
G5A	& involved in glycolysis pathway useful for glucose production	\\ \hline
IGHA1	& immunoglobulin related to the functioning of pancreatic cells	\\ \hline
C8A	& gluconeogenis and in disease state leads to decrease insulin sensitivity	\\ \hline
PIK3C2A	& present in mitochondria and leads helps in energy transfer	\\ \hline
AGER	& involved in glycolytic pathway and in disease state it is down-regulated which leads to less insulin uptake in the body	\\ \hline
NQO1	& is downregulated in obesity and as obesity is directly related to TYD it is directly involved.  	\\ \hline
FBLN1	& It is a marker for arterial extracellular matrix alterations in type 2 diabetes	\\ \hline
CACNA1D	& is downregulated in the disease state which is associated with calcium transport in the cell. 	\\ \hline
EKB	& is a drug molecule associated with the growth factors of a cell and are associated with various cancers	\\ \hline
SLC16A1	& alters lipid metabolism, most notably causing an increase in intracellular triacylglycerol levels. It has been targeted as a novel candidate protein for type 2 diabetes with a possible role in triacylglycerol metabolism.	\\ \hline
CD59	& CD59 impairs complement regulation on erythrocytes leading to poor glycaemic control in diseased sate.	\\ \hline
\end{tabular}
\caption{{\bf Biological functions.} Functional description of proteins having CC=1.}
\label{Table-1}
\end{center}
\end{supptable}

{\bf HPRT1 ( hypoxanthine phosphoribosyltransferase 1)-} Nucleotide salvage enzymes adenylosuccinate synthetase and adenylosuccinate lyase were elevated above normal in the diabetic heart, whereas hypoxanthine-guanine phosphoribosyl transferase was not altered \cite{hprt}.

{\bf Summary-} Housekeeping function.
\\
{\bf NME3 (nucleoside diphosphate kinase 3)-}  Major role in the synthesis of nucleoside triphosphates other than ATP. The ATP gamma phosphate is transferred to the NDP beta phosphate via a ping-pong mechanism, using a phosphorylated active-site intermediate. Probably has a role in normal hematopoiesis by inhibition of granulocyte differentiation and induction of apoptosisn \cite{nme3}.

{\bf Summary-} It is not directly involved in diabetes mellitus II.
\\
{\bf CbX3 (carbenoxolone 3) -} It is shown to improve insulin sensitivity when dosed to healthy human. When it is dosed to lean type 2 diabeteic patient, improved insulin sensitivity, reduced glucose production and glycogenolysis were reported \cite{cbx}.

{\bf Summary-} Involvement in insulin sensitivity, reduced glucose production and glycogenolysis.
\\
{\bf Ddc (dopa decarboxylase)-}It does not have a direct role in occuring DM-II. It is found to Catalyze the decarboxylation of L-3,4-dihydroxyphenylalanine (DOPA) to dopamine, L-5-hydroxytryptophan to serotonin and L-tryptophan to tryptamine \cite{ddc}.

{\bf Summary-}It is not directly involved.
\\
{\bf SERPINA3(Alpha 1-antichymotrypsin)-} In normal cells Alpha 1-antichymotrypsin inhibits the activity of enzymes as proteases, cathepsin G that is found in neutrophils, and chymases. Also serum SERPINA3 levels were higher in T2D which leads to increase in permeability of retinal microvascular endothelial cells, which are involved in the pathogenesis of T2D and/or diabetic retinopathy. Also associated with acute renal allograft rejection \cite{Kalsheker_1996,Klein_1995,Goligorsky_2007,Wellcome_2007}.

{\bf Summary-} This inhibits the activity of enzyme like proteases in neutrophils, and upregulation of its expression level leads to increase in permeability of retinal microvascular leading T2D. And it is also associated with acute renal allograft rejection.
\\
{\bf PSMC5(Proteasome (Prosome, Macropain) 26S Subunit, ATPase, 5)-} In normal cells it is part of Ubiquitin- Proteasome complex which has a major role in targeting cellular proteins for ATP dependent degradation by the 26S. In diseased cells functioning of this complex is either less or absent in which primly leads to T2D \cite{Safia_2011,Rubinsztein_2006,Sakamoto1_2009}.

{\bf Summary-} This protein playes a major role in Ubiquitin- Proteasome complex which is involved in targeting cellular proteins for ATP dependent degradation by the 26S,  downregulation of its expression leads to T2D.
\\
{\bf ANPEP (alanine aminopeptidase):-} In normal cells it is involved in the metabolism of regulatory peptides in renal tubular epithelial cells, and in diabetes is assumed as a primary biomarker due to its increased presence in urine due to Renal tubular dysfunction which leads to T2D or vice versa \cite{Goran_2013}.

{\bf Summary-} This protein acts as a biomarker due to increased concentration in urine due to renal/kidney dysfunction leading T2D.
\\
{\bf PDE3B(Phosphodiesterase 3)-} In normal cells it works in insulin secretion, insulin signaling pathway and fat metabolism. In diseased cells reduced expression of PDE3B in adipose tissue acts as primary event in the development of insulin resistance \cite{Yan_1990}.

{\bf Summary-} This protein playes important role in insulin resistance and insulin signaling pathway, reduced expression leads to T2D.

\section*{Functional properties of proteins denoted as top degree node in disease}

{\bf
UBB(Ubiquitin B)-} It is our 1st top degree node in disease cell line D1 which scored degree of 112 among high degree nodes and bears 444 as its node index. This protein is one of the most conserved proteins known. In normal cell line it scores a degree 40 and bears 1458 as its index no. 
	In normal cells it is part of Ubiquitin- Proteasome complex which has a major role in targeting cellular proteins for ATP dependent degradation by the 26S. In diseased cells functioning of this complex is either less or absent in which directly leads to T2D 
\cite{Safia_2011,Rubinsztein_2006}.
\\
{\bf 
INS (Insulin)-}It is our 1st top degree node in disease cell line D2 which scored degree of 148 among high degree nodes and bears 60 as its node index. In normal cells it scores a degree 118 and bears 111 as its index no and also bears a position of 1st top degree node in normal cell line N3. 
	In normal cells Insulin functions in the normalization of the elevated blood glucose level. But in T2D cells are resistant of insulin and production of insulin gets decreased which directly leads to T2D \cite{Harvey_2007}.

\section*{Functional properties of proteins denoted as top degree node in normal}

{\bf 
UBC(Ubiquitin C)-} It is our 1st top degree node in normal cell line N1 which scored degree of 182 among high degree nodes and bears 1459 as its node index. This protein is one of the most conserved proteins known. In disease cell line it is absent.
	In normal cells it is a part of Ubiquitin- Proteasome complex which has a major role in targeting cellular proteins for ATP dependent degradation by the 26S. In diseased cells functioning of this complex is either less or absent in which primly leads to T2D \cite{Safia_2011,Rubinsztein_2006}.
\\
{\bf ALB(Albumin)-} It is our 2nd node in normal cell line N1 which scored degree of 148 among high degree nodes and bears 414 as its node index. In disease cell lines it has 146 and 79 degree in D1 and D2 respectively. It is as globular, negatively charged, water-soluble protein that is synthesized in the liver. In normal cells functions include maintaining osmotic pressure and transporting a variety of circulating molecules.
 	In diabetes it functions for the same but its production level get slightly elevated in T2D
 \cite{Kelman_1972,Rothschild_1988,Mogensen_1984}.
\\
{\bf CD4(cluster of differentiation 4)-} It is our 3rd node in normal cell line N1 which scored degree of 121 among high degree nodes and bears 1883 as its node index. In disease cell line it is less functional or absent. 
	In normal cells it functions as a co-receptor that assists the T cell receptor (TCR) in communicating with an antigen-presenting cell. CD4 also interacts directly with MHC class II molecules on the surface of the antigen-presenting cell using its extracellular domain \cite{Brady_1993}.
\\
{\bf AKT1(alpha serine/threonine-protein kinas)-} It is our 4th node in normal cell line N1 which scored degree of 108 among high degree nodes and bears 1440 as its node index. In disease cell line it has 309 as its index no.
	This protein is named as RAC-alpha serine/threonine-protein kinase a sub unit of serine/threonine kinase AKT that plays a key in regulating cell survival, insulin signaling in normal cells. But alteration in its activation pathway directly leads to insulin resistance and leads to T2D \cite{ŽDYCHOVÁ_2005,Young_1999,Baohua_2011}.
\\
{\bf
UBA52(Ubiquitin A-52 Residue Ribosomal Protein Fusion Product 1)-} It is our 1st top degree node in Normal cell line N2 which scored degree of 106 among high degree nodes and bears 604 as its node index. This protein is one of the most conserved proteins known. In disease cell line it scores a degree 86 and bears 282 as its index no. 
	In normal cells it is part of Ubiquitin- Proteasome complex which has a major role in targeting cellular proteins for ATP dependent degradation by the 26S. In diseased cells functioning of this complex is either less or absent in which primly leads to T2D. Sometimes Selective expression of UbA52 in the renal tubules suggests that the ubiquitin-proteasome proteolytic system is indeed operative in this compartment of the kidney and might play an important role in diabetic nephropathy. Its release in urine can potentially work as a diagnostic marker \cite{Safia_2011,Rubinsztein_2006,Hassan_2007,MIRCEA_2009}.
\\
{\bf TP53 or P53(Tumor protein p53)-} It is our 2nd  node in normal cell line N2 which scored degree of 105 among high degree nodes and bears 710 as its node index. In disease cell line it is absent.
	In normal cell it encodes the tumor suppressor protein p53, which is known to be involved in cell-cycle control, apoptosis and maintenance of genetic stability, thereby protecting the organism from cellular damage. P53 is expressed in all tissues, but at very low levels under normal conditions \cite{Ventura_2007,Burgdorf_2011}.
\\
{\bf IL6(Interleukin-6)-} It is our 2nd and last  top degree node in Normal cell line N3 which scored degree of 111 among high degree nodes and bears 528 as its node index. In disease cell line it scores a degree 2 and bears 276 as its index no. 
In normal cells it acts as a cytokine and not only involved in inflammation and infection responses but also in the regulation of metabolic, regenerative, signaling, and neural processes. Specific co-relation between T2D and IL-6 is not specified but increased serum levels of interleukin 6 (IL-6) are associated with increased risk of T2D and insulin resistance \cite{Jürgen_2011,Bastard_2000,Kern_2001,Fernandez_2001}.

\section*{Functional properties of disease proteins denoted as top contributing node (D1)}

{\bf PFKL and PFKP(Phosphofructokinase, Liver and platelet)-} In normal cells these are subunits of enzyme called PFK which has key activity in glycolytic pathway and insulin secretion. PFKL and PFKP are isoforms designated as M, P and L for muscle-, platelet-, and liver-type isoforms, respectively. Over expression of PFKL and increased activity of PFK is associated with altered insulin secretion and increased glycolysis rate \cite{Sweet_1995,Dunaway_1987,Elson_1992,Elson1_1994}.

{\bf Summary-} These protein participates in glycolytic pathway and insulin secretion, Their Over expression of leads to alteration in insulin secretion and glycolysis rate.
\\
{\bf COL1A2 and COL3A1(Collagen alpha-2(I) chain and Collagen alpha-1(III) chain)-} In normal cells encodes one of the chains for type I collagen, the fibrillar collagen found in most connective tissues, But in diseased cells due to decreased expressions leads to several heart and lungs related complications 
\cite{Lunteren1_2013,Lunteren_2007}.

{\bf Summary-} These proteins encodes chains for type I collagen in normal cells, But decreased expression levels leads to several heart and lungs related complications.
\\
{\bf H3.3B AND H3F3B(Histone H3) -} In normal cells functions in transcription regulation, DNA repair, DNA replication and chromosomal stability. In diseased cells due to some modifications and alteder expression levels leads to several heart and renal complications \cite{Bhanudas1_2010}.

{\bf Summary-} These proteins participates in transcription regulation, DNA repair, DNA replication and chromosomal stability. Altered expression levels leads to several heart and renal complications.
\\
{\bf PSMD1, PSMD11, PSME1, PSMB8(26S proteasome non-ATPase regulatory subunit 1, 26S proteasome non-ATPase regulatory subunit 11, Proteasome activator complex subunit 1, Proteasome subunit beta type-8)-} These all proteins are subunits of proteasome complex which participates in ubiquitin-proteasome pathway, which works in targeting cellular proteins for ATP dependent protein degradation by 26s. In diseased cells functioning of this complex is either less or absent which leads to obesity and T2D \cite{Aldrin1_2013,Sakamoto1_2009,Mitropoulos_1992,Frayn_1996,Safia_2011}.

{\bf Summary-} These proteins are involved in targeting cellular proteins for ATP dependent protein degradation by 26s. But less or absent expression levels leads to obesity and then T2D. 

\section*{Functional properties of disease proteins denoted as top contributing node (D2)}

{\bf AVPR2(Arginine vasopressin2)-} It is a subunit of Arginine vasopressin (AVP) complex or anti-diuretic hormone, it act as receptor v2 which is responsible for stimulating water retention. Increased expression levels leads to cardio vascular complications, mortality and also limits glucose-induced water loss in T2D 
\cite{Ineke1_2013,Morgenthaler_2010,Bankir_2005}.

{\bf Summary-} This protein act as receptor v2, which is responsible for stimulating water retention. Increased expression levels leads to cardio vascular complications, mortality and also limits glucose-induced water loss in T2D. 
\\
{\bf MT-ND5, MT-ND4, MT-ND1, MT-ND6-} These are subunits of respiratory complex 1 (NADH: ubiquinone oxidoreductase) encoded by mitochondria. They participates in oxidative phosphorylation, ion exchange, cell signaling, insulin signaling and insulin resistance. Any alteration in mitochondrial genome leads to T2D 
\cite{Michal1_2007,Kaasik_2001,Hood_2001,Saltiel_2000,Sheperd_1999,Shulman_2000,Vladimir_2005}.

{\bf Summary-} These proteins are involved in insulin signaling and insulin resistance. Alteration in any of these proteins genome leads to T2D.
\\
{\bf MT-CO3(Mitochondrially Encoded Cytochrome C Oxidase III)-} This protein functions as an enzyme complex which playes vital roles in body weight regulation, in diseased cells due to decreased expression levels leads to mitochondrial dysfunction, obesity, insulin resistance, and T2D
 \cite{Hutchison_2012,Gastaldi_2007,Short_2003,Holmstrom_2012,Chung_2012}.

{\bf Summary-} This protein functions as an enzyme complex which playes vital roles in body weight regulation, in diseased cells due to decreased expression levels leads to mitochondrial dysfunction, obesity, insulin resistance, and T2D.
\\
{\bf SEL1L(Suppressor Of Lin-12-Like) and SELS/VIMP (Selenoprotein S/VCP-Interacting Membrane Protein)-} They are ER(endoplasmic reticulum) membrane protein, which functions for retrotranslocation of misfolded proteins from the endoplasmic reticulum in normal cells, but in diseased cells a deficiency of SEL1L and SELS causes systemic ER stress leading retarded glucose-stimulated insulin secretion leading T2D 
\cite{Adam1_2011,Amanda_2013}.

{\bf Summary-} These proteins functions in retrotranslocation of misfolded proteins, Their deficiency leads systemic ER stress leading retarded glucose-stimulated insulin secretion leading T2D.
\\
{\bf CALM1, CALM2, CALM3(Calmodulin 1,2,3)-} These are members of the EF-hand calcium-binding protein family (calmodulin) which plays vital role in Calcium/Calmodulin-Dependent Protein Kinase II pathway responsible in insulin signaling. Their decreased expression levels leads to insulin resistance and obesity causing T2D \cite{Ozcan1_2013}.

{\bf Summary-} These proteins functions in Calcium/Calmodulin-Dependent Protein Kinase II pathway. Their decreased expression levels leads to insulin resistance and obesity causing T2D.
\\
{\bf HLA-DQA1(Major Histocompatibility Complex, Class II, DQ Alpha 1)-} It plays a central role in the immune system by presenting peptides derived from extracellular proteins. Class II molecules are expressed in antigen-presenting cells (APC: B lymphocytes, dendritic cells, macrophages). In diseased cells its increased expressions subsequently increases susceptibility for T2D \cite{Ze-Jun1_2013,Entrez4}.

{\bf Summary-} This protein plays important role in immune system, and in diseased cells its increased expressions subsequently increases susceptibility for T2D.
\\
{\bf C1S(Complement Component 1, S Subcomponent)-} It is a serine protease that combines with C1q and C1r to form C1. It plays an important role in the initiation of inflammatory process, in diseased cells Upregulation of C1s expression in adipose cells leads obesity and insulin resistant in T2D \cite{Jinhui1_2007}.

{\bf Summary-} This protein is responsible for initiation of inflammatory process, upregulation in its expression leads to obesity and insulin resistant.
\\
{\bf YWHAE(tyrosine 3-monooxygenase/tryptophan 5-monooxygenase activation protein, epsilon)-} In normal cells it functions in  signal transduction by binding to phosphoserine-containing proteins in insulin signaling and sensitivity, alteration in its expression levels leads to obesity and T2D \cite{Adeyemo1_2009}.

{\bf Summary-} This protein functions in signal transduction and alteration in expression levels causes defect in insulin secretion and sensitivity and obesity.

\end{document}